\documentclass{IEEEtran4PSCC}
\ifCLASSINFOpdf
   \usepackage[pdftex]{graphicx}
\else
   \usepackage[dvips]{graphicx}
\fi
\usepackage[cmex10]{amsmath}

\usepackage[utf8]{inputenc}
\usepackage[pdftex]{graphicx}
\usepackage[cmex10]{amsmath}
\usepackage{amsfonts}
\usepackage{subcaption}
\usepackage{comment}
\usepackage{float}
\usepackage[dvipsnames]{xcolor}
\usepackage[hyphens]{url}
\usepackage{hyperref}

\usepackage{comment}
\usepackage{mathtools}
\usepackage[english]{babel}
\usepackage{amsthm}
\usepackage{amssymb,amsmath,amsthm}

\newtheorem*{remark}{Remark}
\usepackage{graphicx}
\usepackage{svg}
\usepackage[pdftex]{graphicx}
\usepackage{epstopdf}

\let\norm\undefined 
\DeclarePairedDelimiter\norm{\lVert}{\rVert}

\makeatletter
\let\old@ps@headings\ps@headings
\let\old@ps@IEEEtitlepagestyle\ps@IEEEtitlepagestyle
\def\psccfooter#1{%
    \def\ps@headings{%
        \old@ps@headings%
        \def\@oddfoot{\strut\hfill#1\hfill\strut}%
        \def\@evenfoot{\strut\hfill#1\hfill\strut}%
    }%
    \def\ps@IEEEtitlepagestyle{%
        \old@ps@IEEEtitlepagestyle%
        \def\@oddfoot{\strut\hfill#1\hfill\strut}%
        \def\@evenfoot{\strut\hfill#1\hfill\strut}%
    }%
    \ps@headings%
}
\makeatother

\psccfooter{%
        \parbox{\textwidth}{\hrulefill \\ \small{23rd Power Systems Computation Conference} \hfill \begin{minipage}{0.2\textwidth}\centering \vspace*{4pt} \includegraphics[scale=0.06]{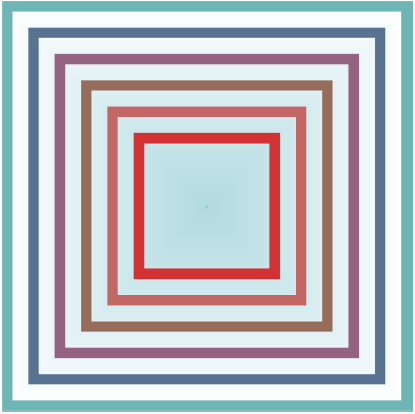}\\\small{PSCC 2024} \end{minipage} \hfill \small{Paris, France --- June 4 -- 7, 2024}}%
}

\begin{document}
%
\title{Transmission line dynamics on inverter-dominated grids: analysis and simulations}

\author{

\IEEEauthorblockN{Gabriel E. Col\'on-Reyes$^\star$, Ruth Kravis$^\star$, Sunash B. Sharma$^\star$, Duncan S. Callaway$^{\star, \dagger}$}
\IEEEauthorblockA{$^\star$Department of Electrical Engineering and Computer Sciences, University of California, Berkeley, United States \\
$^\dagger$Energy and Resources Group, University of California, Berkeley, United States\\
\{gecolonr, ruth.kravis, sbsharma, dcal\}@berkeley.edu}}


\maketitle

\begin{abstract}
In this work, we study fast time scale dynamic interactions between inverters, synchronous machines, and transmission lines. The overlapping time scales between inverter controls and electromagnetic phenomena idfentified in recent years has necessitated a re-evaluation of assumptions made in power system dynamics studies. We utilize an open-source modeling platform to perform both small signal stability and dynamic time domain studies of networks containing inverters, machines, and loads. We use transmission line models of varying fidelity, including models with multiple segments and frequency dependence. Our results indicate that, for the cases we study, line dynamics are not important for characterizing small signal stability.  However, while in many cases the high fidelity line models are unnecessary for dynamic simulations, there are some cases in which simpler models omit dynamics that could be important to the operation of inner inverter control loops. 
\end{abstract} 
\begin{IEEEkeywords}
Low-inertia grids, power system dynamics, power system stability, transmission line dynamics.
\end{IEEEkeywords}

\thanksto{\noindent Submitted to the 23rd Power Systems Computation Conference (PSCC 2024).  This research was supported by the U.S. Department of Energy's Solar Energy Technologies Office through award 38637 (UNIFI Consortium).}

\section{Introduction}

This paper addresses the question of whether or not electromagnetic (EM) dynamics should be included in power system dynamic interaction studies.  These dynamics have been largely ignored due to the natural time scale separation between synchronous machine (SM) states and EM dynamics.  However, because converter-interfaced generation (CIG) sources have feedback control loops whose time constants overlap with EM dynamics, it is no longer clear whether fast time scale dynamics can be neglected~\cite{ENTSOe_interaction}.  This question has become increasingly prominent with the functionality of grid-forming inverters (GFM) to regulate frequency and voltage magnitude.  Addressing this question has important compuational implications:  whereas phasor-domain modeling tools that ignore fast time scale dynamics can simulate power system dynamics in a matter of seconds or less, electromagnetic transient (EMT) software tools can require many hours to simulate even relatively small systems \cite{kenyon2021validation}. 


The central component in this question – and what determines the difference between phasor domain simulations and EMT simulations – are the transmission lines (TLs).  Phasor domain simulations typically have detailed load, machine and inverter models that are coupled via algebraic line models that ignore the EM dynamics.  On the other hand, EMT methods model the detailed physics of TLs – including wave propagation and frequency dependence – in tandem with other power system elements.  This enables EM time scale phenomena from a variety of system components to accurately propagate from one network location to another.   

Several recent studies have found that simple models of TL dynamics can impact small signal stability classifications~\cite{markovic2021understanding, henriquez2020grid, Gros_Colombino_Brouillon_Dorfler_2019}, in some situations showing that line dynamics can destabilize grids with high shares of CIG, while stabilizing them in others.  In these papers, TLs are modeled with a single RL branch~\cite{markovic2021understanding,Gros_Colombino_Brouillon_Dorfler_2019} or dynamic lumped-parameter $\pi$ models~\cite{henriquez2020grid}.  While these models provide higher fidelity dynamics than algebraic models, they do not capture transmission line frequency dependence or wave propagation dynamics associated with distributed parameter or multi-segment line models.  However, research on direct current line models that interface with power electronic converters \cite{D’Arco_Beerten_Suul_2015_cable_MOR, D’Arco_Suul_Beerten_2019, D’Arco_Suul_Beerten_2021_config} has shown that frequency dependence and multi-segment dynamics can produce qualitatively different dynamics relative to single segment frequency independent models.



This paper aims to contribute to the discussion around TL modeling choice in system-level dynamics and stability by examining the impact of higher fidelity TL models on simulation outcomes.  In addition to comparing algebraic network models to dynamic $\pi$ models, we examine multi-segment TL models with and without frequency dependent dynamics.  Our contributions are as follows:

\begin{itemize}
	\item First, building on the open-source simulation package \texttt{PowerSimulationsDynamics.jl} \cite{lara2023powersimulationsdynamicsjl}, we construct a TL modeling package that allows for comparison among several transmission line models. This facilitates comparisons between a large variety of TL line modeling choices in a single, fast-to-simulate, open-source simulation package, rather than relying on aligning simulation assumptions between commercial phasor-domain simulation software and EMT simulation software.  
	\item Second, we examine the effect of TL model choice on the small signal stability of power systems with a mix of grid-forming CIG and SMs.  For the GFM-SM cases we study, adding TL fidelity with more segments and frequency dependence alters both the real and imaginary parts of the system's eigenvalues.  However, we find that the loading and line length at which systems become unstable is independent of TL model.  These results are consistent with a subset of results from earlier studies and provide additional detail with respect to the effects of including higher fidelity line models. 
	\item Third, we examine the effect of TL model choice on dynamic simulation outcomes for branch trips across several case studies, loading scenarios, and line lengths, for systems with a mix of GFM and SMs.  We observe various differences in voltage and current dynamics; however, these differences do not produce sustained dynamics. For a subset of experiments, higher fidelity models produce transients of significantly different amplitude and direction than the lower fidelity models -- suggesting that simulations with low order line models may miss important dynamic events, especially in the study of the action of CIG inner control loops.  
\end{itemize} 

\textit{Notation: }Dot notation indicates the time derivative of a variable, i.e., $\dot{x} = \frac{dx}{dt}$. Bold lower-case symbols are used to represent complex variables in the $dq$ or $RI$ reference frames, e.g. $\boldsymbol{x}=x_R + jx_I$. $\lambda_A \in \sigma(A)$ is an eigenvalue of $A$, where $\sigma(A)$ is the spectrum of $A \in \mathbb{R}^{n \times n}$. $\norm{\cdot}$ is the Euclidean vector norm.

\section{Modeling}

\subsection{EMT and positive sequence simulations}

EMT tools such as PSCAD simulate systems via a sequence of coupled nodal equations in which TL dynamics and bus-level device dynamics are represented in a discrete time formulation.  This  allows line wave propagation dynamics and delays to be captured with high fidelity.  In contrast, phasor domain tools such as PSS/e and PSLF solve the dynamics of devices at each bus and, in order to accelerate simulation times, represent the network model as an algebraic system solved in a separate power flow step.  Phasor domain tools are further accelerated relative to EMT tools by using ordinary differential equation (ODE) solvers that numerically integrate dynamics far faster than discrete time EMT solvers \cite{revisiting_methods_lara}.  

These differences complicate simulation-based comparisons of dynamic phenomena with and without TL dynamics. EMT software tools are slow to solve, and significant modeling effort (parameter choice, model harmonization) is required to ensure that simulation outputs without line dynamics match industry standard phasor domain modeling tools. Moreover, Bergeron-style discrete-time formulations of high-frequency dynamics for TLs produce delay-\textit{difference} equations that cannot be easily deployed in a small signal stability analysis. 

In this paper we leverage the Julia-based modeling and simulation package \texttt{PowerSimulationsDynamics.jl} (PSID).  PSID can simulate power system dynamics in the phasor domain as well as in a balanced $dq$ form, and it is capable of precisely reproducing the output of PSS/e in the phasor domain, even on very large network models, as well as PSCAD EMT with fast inverter and machine dynamics \cite{lara2023powersimulationsdynamicsjl}. In contrast to PSCAD, because PSID preserves a $dq$ ODE formulation of the system model, it can leverage a suite of numerical integration solvers that enable significantly faster simulation speeds of power system dynamics.  

However, at the time of writing this artcile, PSID has only been developed to model dynamic $\pi$ representations of TL dynamics.  In this paper we extend PSID to approximate distributed parameter TL dynamics with and without frequency dependence.  We do so by adapting a multi-segment multi-branch modeling approach \cite{Beerten_D’Arco_Suul_2016, D’Arco_Suul_Beerten_2021_config} originally developed for modeling direct current cables, to work in PSID's multi-machine AC power system simulation setting. 

\subsection{Power System Components}
\subsubsection{Generator}

In PSID, a generator is composed of five main components: a stator, a shaft, a turbine governor, a power system stabilizer (PSS), and an automatic voltage regulator (AVR). Our choice of models for each of the components is based on the time scales over which the dynamics of such component evolves relative to our interest in fast time scales of inverter controls and TL dynamics.

The models that we choose for each of the components of the generator, along with the corresponding choices for parameters, can be found in Chapters 15 and 16 of \cite{milano2010power}. We adopt the six-state Anderson-Fouad machine stator model. This model incorporates transient and subtransient EM stator dynamics. We choose a shaft model given by the swing equations with damping. Further, we opt for a fixed input turbine governor because its output will be relatively constant in the time scales of interest. Because of its use for slow time scales, and thus its irrelevance to our analysis, we choose not to include a PSS. Lastly, we opt for a Type 1 AVR  to capture voltage control dynamics.

\subsubsection{Inverter}

In PSID, an inverter is composed of six main components: a DC voltage source, a model for the switches, an output inverter filter, an outer grid-forming (GFM) or grid-following (GFL) control loop, an inner control loop, and a frequency estimator.

The models we choose for these components and its parameters come from \cite{Beerten_D’Arco_Suul_2016}. We choose a fixed DC voltage source model, an LCL passive filter, a virtual synchronous machine GFM model for the outer loops, nested proportional-integral (PI) loops for the inner control loops, and a phase locked loop for damping of the virtual frequency. We choose an averaged model for the switches \cite{yazdani2010voltage}. For some experiments, we choose to have the inverter be a GFL source, modeled by active and reactive power PI controllers \cite{kenyon2021open}.

\begin{figure*}
 \begin{subfigure}{.6\columnwidth}
    \includegraphics[width=.8\columnwidth]{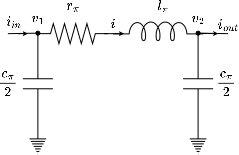}
    \caption{}
        \label{fig:pi_schematic}
  \end{subfigure}
    \begin{subfigure}{1.4\columnwidth}
        \centering
        \includegraphics[width=\columnwidth]{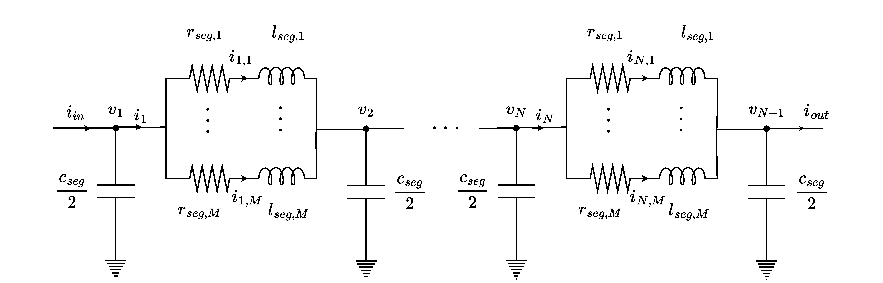}
        \caption{}
        \label{fig:ms_schematic}
    \end{subfigure}
    \caption{Line model schematics: (a) $\pi$ model ($statpi$ and $dynpi$), and (b) Multi-segment model ($MSSB$ and $MSMB$).}
    \label{fig:line_models}
\end{figure*}

\subsection{Transmission lines}\label{sec:tx_models}
We consider TL models that can be expressed as linear time-invariant state-space models, since these are compatible with differential-algebraic equation (DAE) representations of other system components.

To model TLs of arbitrary length (denoted $\ell$), we use per unit length parameters for impedance, $z_{km} = r_{km} + j\omega l_{km}$, and susceptance, $b_{km} = g_{km} + j\omega c_{km}$. At a particular operating frequency $\omega$, we compute the lumped parameter equivalent $\pi$ model according to the following equations, which include hyperbolic correction factors from the steady state solution to Telegrapher's equations:
\begin{align}
    z_{\pi} &= z_{km} \ell \left( \frac{\sinh(\gamma \ell)}{\gamma \ell} \right) \label{eqn:hyperbolic1}\\ 
    y_{\pi} &= y_{km} \ell \left( \frac{\tanh(\gamma \ell/2)}{\gamma \ell/2} \right) \label{eqn:hyperbolic2}\\
    \gamma &= \sqrt{z_{km}y_{km}} \label{eqn:gamma}
\end{align}
where $\ell$ is line length, as distinct from $l$ which refers to inductance.
We assume $g_{km}=0$, which does not imply that $g_\pi = \mathrm{Re}(y_{\pi})=0$. However, we choose to overwrite $g_\pi = 0$. From this model, we can compute equivalent $r, l, c$ as $r_\pi=\mathrm{Re}(z_{\pi}), l_\pi= x_\pi/\omega = \mathrm{Im}(z_{\pi})/\omega, c_\pi= b_\pi/\omega = \mathrm{Im}(y_{\pi})/\omega$. 

\subsubsection{Algebraic $\pi$ model (statpi)}

The algebraic $\pi$ model has the form shown in Fig.~\ref{fig:pi_schematic}. It assumes that any line dynamics are stable and settle quickly compared to other system dynamics. Therefore, the differential terms associated with the line capacitance and inductance are set to zero, giving an algebraic model as follows: 
\begin{align}
    \boldsymbol{i_{in}} &= \left(\frac{1}{z_\pi} + y_\pi\right)\boldsymbol{v_1} - \frac{1}{z_\pi}\boldsymbol{v_2} \\
    \boldsymbol{i_{out}} &= -\frac{1}{z_\pi}\boldsymbol{v_1} - \left(\frac{1}{z_\pi}+y_\pi\right)\boldsymbol{v_2} \label{eqn:statpi}
\end{align}
Since this model is purely algebraic, any dynamics on $\boldsymbol{i_1}$, $\boldsymbol{i_2}$, $\boldsymbol{v_1}$, and $\boldsymbol{v_2}$ that arise due to the interconnection of the line with dynamic devices will instantaneously appear at the other end of the line. 

\subsubsection{Dynamic $\pi$ model (dynpi)}
The dynamic $\pi$ model has the same structure as shown in Fig.~\ref{fig:pi_schematic}, but it includes dynamics on line current and voltage states:
\begin{align}
    \frac{l_{\pi}}{\omega_b
}\frac{d\boldsymbol{i}}{dt} &= (\boldsymbol{v_1} - \boldsymbol{v_2}) - z_\pi\boldsymbol{i} \\
    \frac{c_{\pi}}{2
} \frac{1}{\omega_b}\frac{d\boldsymbol{v_1}}{dt} &= (\boldsymbol{i_{in}} - \boldsymbol{i}) - y_\pi\boldsymbol{v_1} \\
    \frac{c_{\pi}}{2}\frac{1}{\omega_b}\frac{d\boldsymbol{v_2}}{dt} &= (\boldsymbol{i} - \boldsymbol{i_{out}}) - y_\pi \boldsymbol{v_2} \label{eqn:dynpi} 
\end{align}

Similar to $statpi$, $dynpi$ only captures the effect of the distributed line parameters in steady state.

\subsubsection{Multi-segment single-branch $\pi$ model ($MSSB$)}

In what follows, a line with multiple ``segments'' is one that is divided into a discrete set of identical length $\pi$ components connected in series.  In addition, a line with multiple ``branches'' is one in which each segment is divided into a set of parallel branches, each with different impedances, to capture frequency dependent line characteristics.

The multi-segment single branch $\pi$ model ($MSSB$) consists of $N$ $\pi$-shaped segments.  The parameters for each segment are given by:
\begin{align}
    r_{seg} &= r_{km}\ell_{seg} \\
    l_{seg} &= l_{km}\ell_{seg} \\
    c_{seg} &= c_{km}\ell_{seg} 
\end{align}
where $\ell_{seg} = \frac{\ell}{N}$. Further, $z_{seg} = r_{seg} + j\omega l_{seg}$ and $
y_{seg} = j\omega c_{seg}$. This model is seen in Fig. \ref{fig:ms_schematic} with $M = 1$, namely a single RL branch for each segment. The $i^{th}$ segment of an $N$-segment $MSSB$ model is defined by the following equations: 
\begin{align}
    \frac{l_{seg}}{\omega_b
}\frac{d\boldsymbol{i_{i}}}{dt} &= (\boldsymbol{v_i} - \boldsymbol{v_{i+1}}) - z_{seg}\boldsymbol{i_{i}} \\
    \frac{c_{seg}}{2}\frac{1}{\omega_b}\frac{d\boldsymbol{v_i}}{dt} &= (\boldsymbol{i_{i-1}} - \boldsymbol{i_{i}}) - y_{seg}\boldsymbol{v_i} \\
    \frac{c_{seg}}{2}\frac{1}{\omega_b}\frac{d\boldsymbol{v_{i+1}}}{dt} &= (\boldsymbol{i_{i}} - \boldsymbol{i_{i+1}}) - y_{seg}\boldsymbol{v_{i+1}} \label{eqn:mssb} 
\end{align}

Note that for $i=1$, $i_{i-1}=i_{in}$, and for $i=N$, $i_{i+1}=i_{out}$. As $N$ is increased, the $MSSB$ model more closely approximates the equivalent $\pi$ model in steady state frequency response. The advantage of explicitly representing segments is that unlike the equivalent $\pi$ model, it captures the distributed nature of the line parameters in both transient and steady state responses. 

\subsubsection{Frequency dependent multi-segment multi-branch $\pi$ model ($MSMB$)}

The $MSMB$ model has $M$ branch currents per segment. Let the subscript $m=1,...,M$ denote the $m^{th}$ parallel branch. Therefore, the equations for the $i^{th}$ segment are:
\begin{align}
    \frac{l_{seg,m}}{\omega_b
}\frac{d\boldsymbol{i_{i,m}}}{dt} &= (\boldsymbol{v_i} - \boldsymbol{v_{i+1}}) - z_{seg,m}\boldsymbol{i_{i,m}} \quad \forall m\\
    \frac{c_{seg}}{2}\frac{1}{\omega_b}\frac{d\boldsymbol{v_i}}{dt} &= (\boldsymbol{i_{i-1}} - \boldsymbol{i_{i}}) - y_{seg}\boldsymbol{v_i} \\
    \frac{c_{seg}}{2}\frac{1}{\omega_b}\frac{d\boldsymbol{v_{i+1}}}{dt} &= (\boldsymbol{i_{i}} - \boldsymbol{i_{i+1}}) - y_{seg}\boldsymbol{v_{i+1}} \label{eqn:mssb} 
\end{align}
where $z_{seg,m} = r_{seg,m} + j\omega l_{seg,m}$. Note that $\boldsymbol{i_i} = \sum_{m=1}^{M}{\boldsymbol{i_{i,m}}}$ in the voltage equations, as seen in Fig. \ref{fig:ms_schematic}.

The $MSMB$ is considered the highest fidelity line model of those presented, since it captures dynamics along the line's length as well as the additional damping that arises due to the frequency dependence of the line parameters \cite{D’Arco_Suul_Beerten_2021_config}.

\subsection{Line parameters}

For the $MSMB$ model, we obtain $r_{km,m}$ and $l_{km,m}$ for each branch from known line data via vector fitting~\cite{gustavsen1999rational, Beerten_D’Arco_Suul_2016}, a form of parameter estimation based on real transmission line data. In this paper, we use the frequency dependent line impedance parameters from~\cite{Dommel_1985} as a starting point\footnote{We use the data in Table 3.}.  We derive parameters for all other line models from the $MSMB$ model by finding the equivalent parallelized branch impedance at nominal frequency for the $MSMB$ model. This yields the $MSSB$ $z_{km}$, which we use in Equations~\ref{eqn:hyperbolic1}, \ref{eqn:hyperbolic2} and~\ref{eqn:gamma} to obtain the lumped parameter model parameters for $statpi$ and $dynpi$. It proves important to follow a procedure like this to compute impedances and susceptances for all models to ensure that at the steady state operating frequency all lines have the same impedance, resulting in the same network power flow solution. 

\subsection{Aggregate system model}

All the models described above can be written in DAE form. Linking the devices and network, we arrive at a mathematical form as follows:
\begin{equation}
    \begin{bmatrix}
    \dot{x} \\ 0
\end{bmatrix}
= 
\begin{bmatrix}
    f(x, y, u) \\ g(x, y, u)
\end{bmatrix}
\end{equation}
Here, $x \in \mathbb{R} ^n$ are the system's dynamic states, $y \in \mathbb{R} ^m$ are the system's algebraic states, $u \in \mathbb{R} ^p$ are the system inputs. $f:\mathbb{R}^n \times \mathbb{R}^m \times \mathbb{R}^p \rightarrow \mathbb{R}^n$ and $g:\mathbb{R}^n \times \mathbb{R}^m \times \mathbb{R}^p \rightarrow \mathbb{R}^m$ are the vector equations associated to the dynamics of the network and the algebraic constraints. 

We are interested in studying the stability of such a system, in particular, its small signal stability. In general, $f$ is a nonlinear vector field. Therefore, to study the small signal stability of the system we find an equilibrium point $(x^\star, y^\star, 0)$ by setting $\dot{x} = 0$ and solving the nonlinear system of equations, linearize them around that point, and arrive at a set of linear dynamics that characterize the behavior of the system in the vicinity of that equilibrium point. The resulting equations will then be of the following form:
\begin{equation}
    \Delta\dot{x} = J(x^\star, y^\star) \Delta x
\end{equation} 

Here, $J(x^\star, y^\star)$ is the reduced system Jacobian matrix. By studying the eigenvalues of this matrix associated to the linearization, we can determine if $(x^\star, y^\star, 0)$ is a stable or unstable operating condition for the network. See \cite{henriquez2020grid} or \cite{lara2023powersimulationsdynamicsjl} for further details on the linearization process in PSID.

\section{Test Cases}
To investigate dynamic interactions under the different line models presented, we choose a simplified two bus test case, and the IEEE 9 Bus test case.

\begin{remark}
    We choose not to use an infinite bus in any of our studies because it would instantaneously produce or consume whatever real and reactive power are necessary to maintain its voltage, which has an unrealistic effect on the dynamic results. Instead, we choose a SM/GFM inverter as the voltage angle reference bus to solve the initializing power flow problem and subsequently simulate dynamics of the multi-machine system. 
\end{remark}

\subsection{Two Bus test case}
\begin{figure}[h]
    \centering
\includegraphics[width=.5\columnwidth]{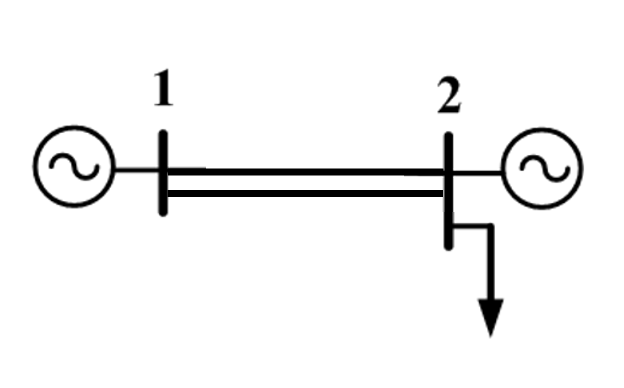}
    \caption{Two Bus test system single line diagram.}
    \label{fig:twobus_schematic}
\end{figure}

The two bus test case, shown in Fig.~\ref{fig:twobus_schematic}, has two generation sources connected by two identical transmission lines in parallel. By default, $\ell=100$ km. 

The test case has a constant impedance load, which can be located at either one of the two buses. We choose $p_{load}$ and $q_{load}$ according to the line's surge impedance loading $(SIL) = \frac{V^2}{Z_c},$
where $V$ is the nominal system voltage (230 kV) and $Z_c = \sqrt{z_{km}/y_{km}}$ is the line's characteristic impedance. Normally, $SIL$ is computed for lossless lines, however here we consider losses to give nominal $SIL$ values for real and reactive power. For our parameters of $V$, $z_{km}$, and $y_{km}$, we arrive at $p_{load} = 2.05$ p.u. and $q_{load} = 0.08$ p.u.

\subsection{IEEE 9 Bus test case}

\begin{figure}[h]
    \centering\includegraphics[width=.35\textwidth]{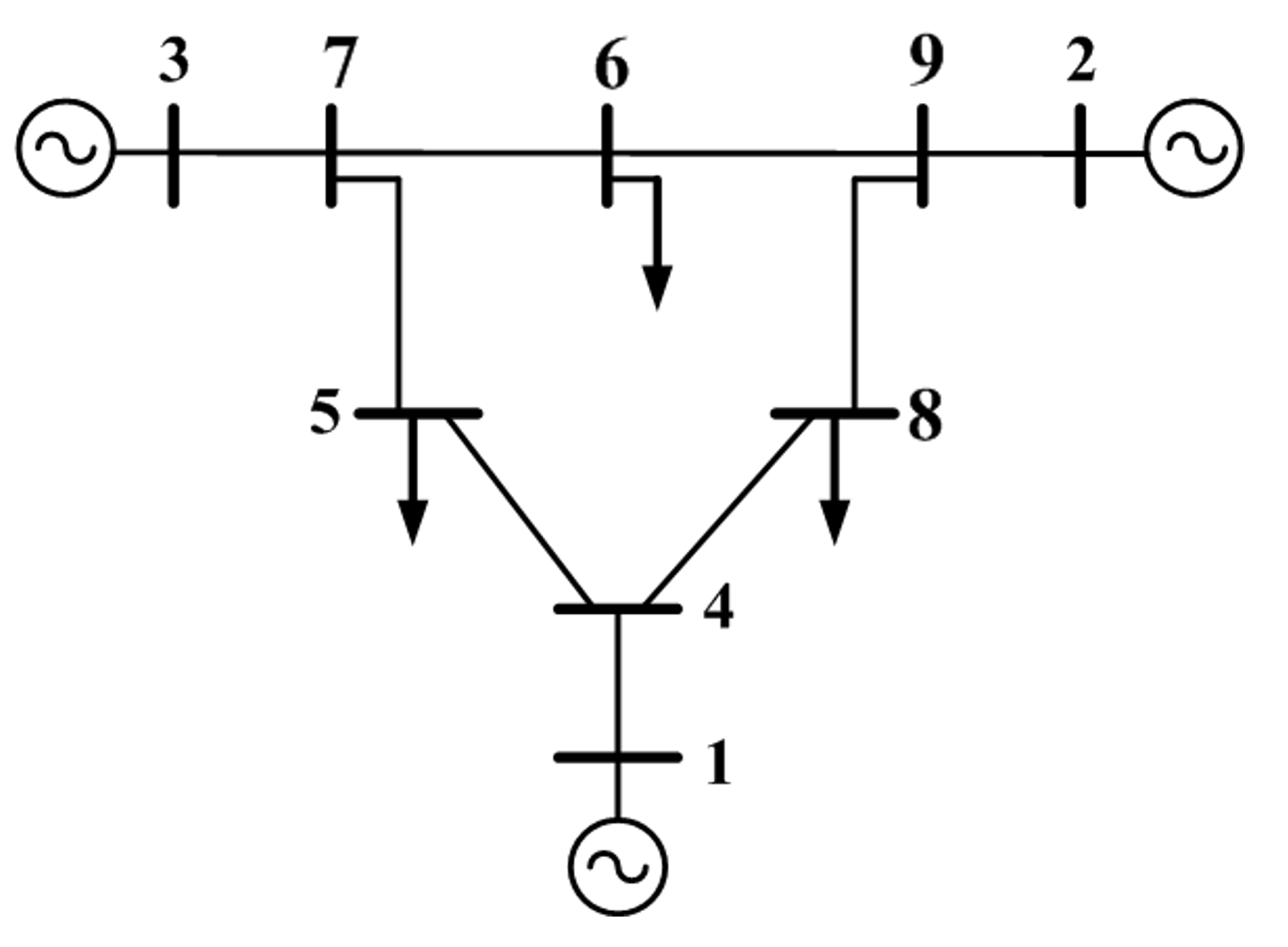}
    \caption{IEEE 9 Bus test case single line diagram \cite{9bus_figure}}.
    \label{figs:9bus}
\end{figure}

We also study the IEEE 9 Bus test case, shown in Fig. \ref{figs:9bus}. We place two identical GFM inverters at buses 1 and 3, and a SM at bus 2. The test case also includes three constant impedance loads at buses 5, 6 and 8. We set nominal loading and line lengths according to \cite{Berard2017}. 

\subsection{Experiment types}
We construct different experiments by varying loading and line lengths. We vary loading because it is a key variable of interest for a system operator identifying safe operating limits of a system. We vary line lengths under the hypothesis that differences between line models may be revealed only for longer lines. 

We vary line length by scaling all nominal line lengths by a \textit{line scale}. We vary all loads' real and reactive power consumption by multiplying nominal values by a \textit{load scale}, as well as all generators' real and reactive power setpoints.

\begin{remark}
    We choose the default load for the two bus test case as the $SIL$ for the line, however the IEEE 9 Bus test case has predefined loads, all of which are different than the line $SIL$. Therefore, a particular load scale factor scales a different `base' load in the IEEE 9 Bus test case compared to the Two Bus test case. 
\end{remark}

\section{Small signal analysis}


We compare small signal stability under our four different line models. Following the approach developed in \cite{D’Arco_Beerten_Suul_2015_cable_MOR}, we choose the segment length for $MSSB$ and $MSMB$ by selecting $N$ such that the highest frequency oscillation of the line model is at least as large as the highest frequency oscillation mode of the GFM and SM. For the $MSMB$ model we choose three parallel branches (i.e., $M=3$), which is consistent with the modelling recommendations in \cite{D’Arco_Suul_Beerten_2021_config}. 

\subsection{Two Bus test case}

\subsubsection{Stability boundary}
\begin{figure}[h]
 \begin{subfigure}{0.48\columnwidth}
    \includegraphics[width=\columnwidth]{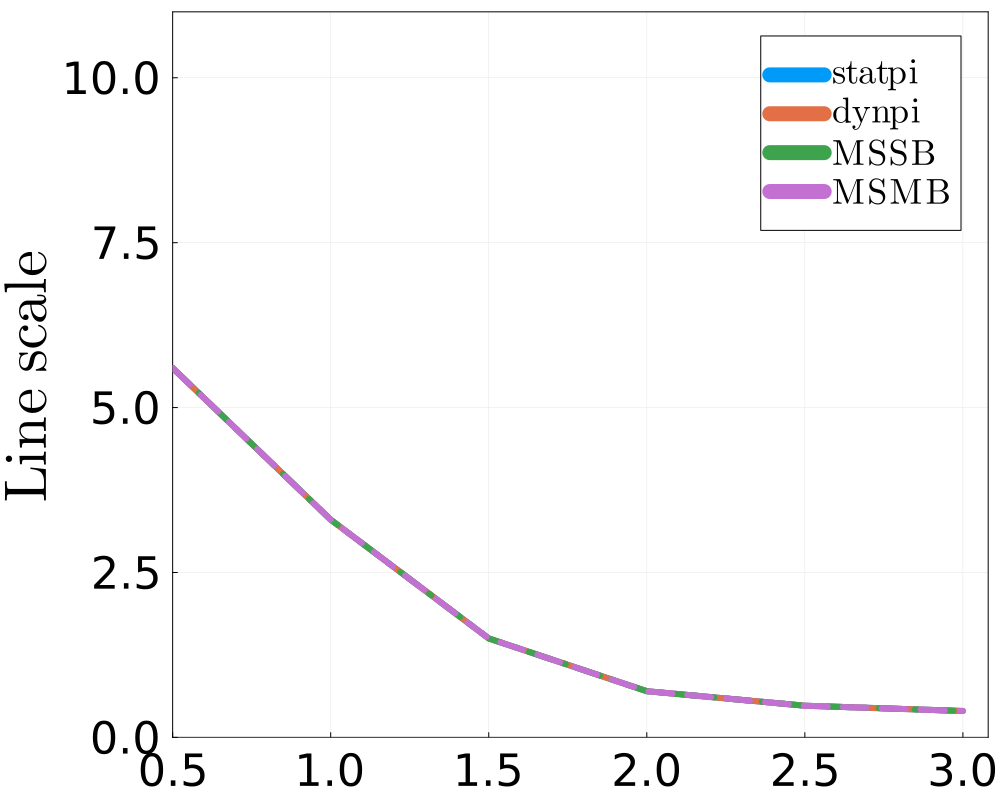}
        \caption{}
        \label{fig:inv_mach_load_inv}
  \end{subfigure}
    \begin{subfigure}{0.48\columnwidth}
        \centering
        \includegraphics[width=\columnwidth]{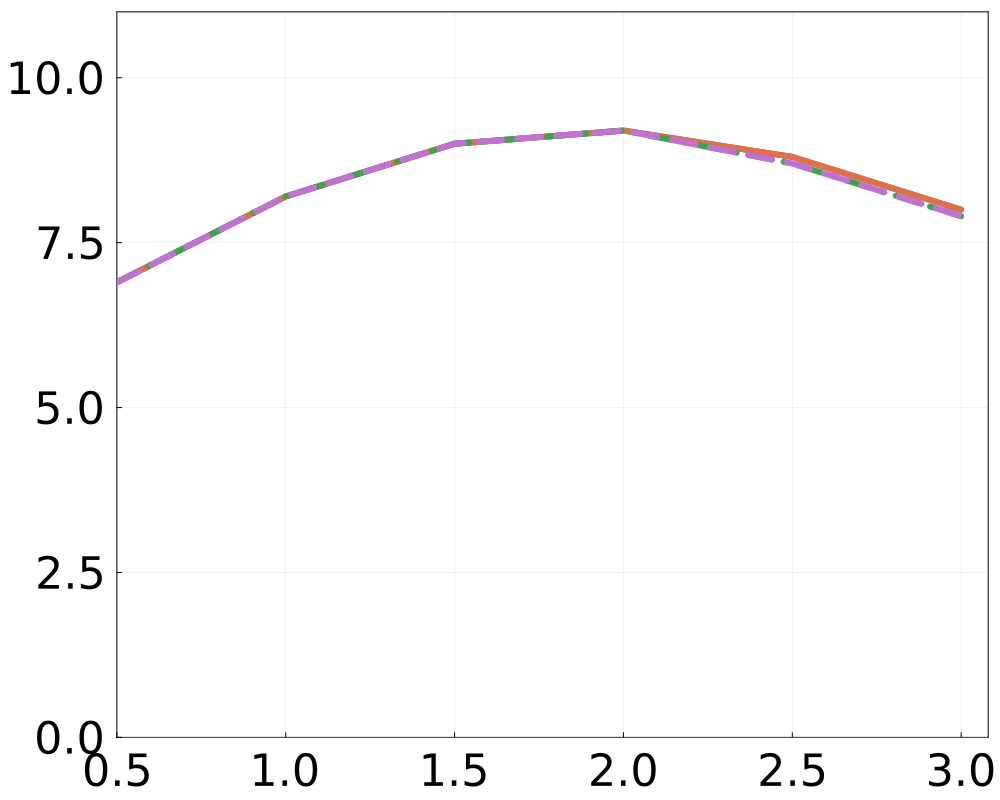}
        \caption{}
        \label{fig:inv_mach_load_mach}
    \end{subfigure}
    \vfill
\begin{subfigure}{0.48\columnwidth}
    \includegraphics[width=\columnwidth]{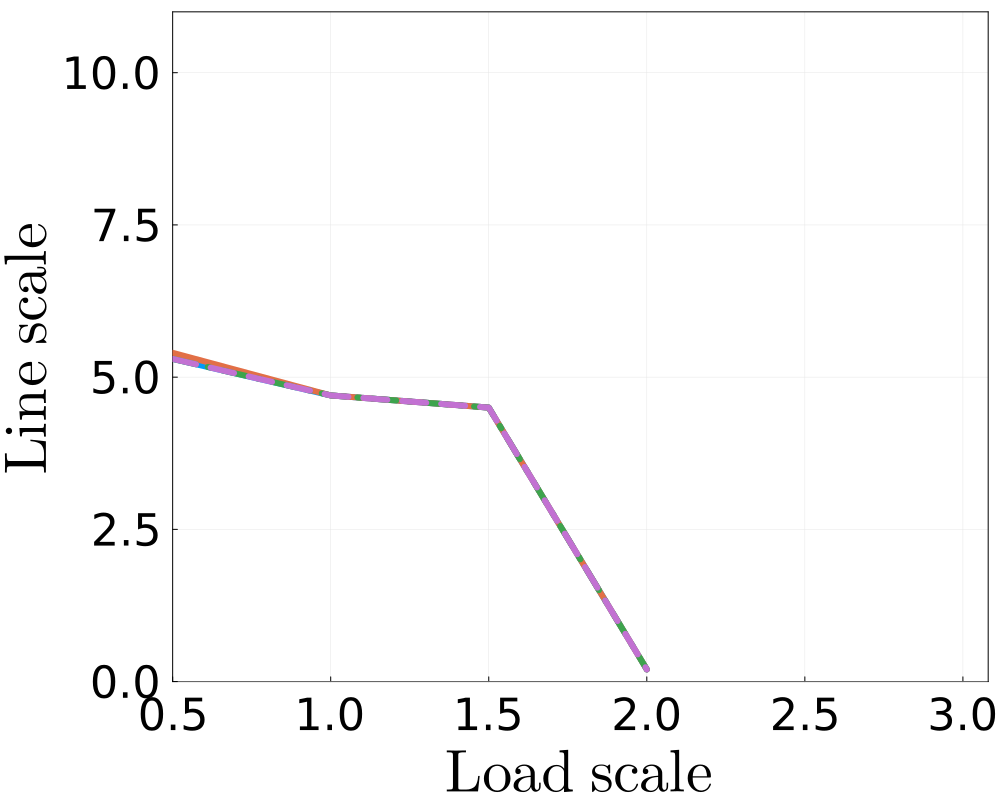}
        \caption{}
        \label{fig:mach_mach}
  \end{subfigure}
    \begin{subfigure}{0.48\columnwidth}
        \centering
    \includegraphics[width=\textwidth]{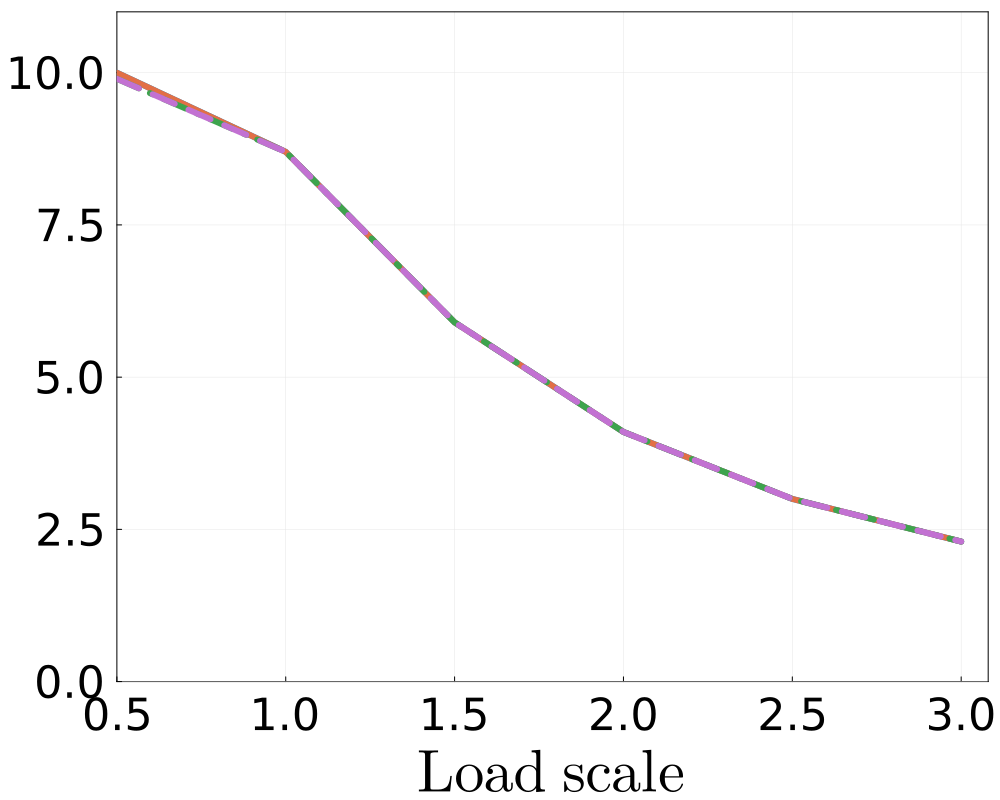}
        \caption{}
        \label{fig:inv_inv}
    \end{subfigure}
    \caption{Line length where the system loses stability on the $y$-axis (measured in terms of $line \ scale$) as a function of system loading on the $x$-axis (measured in terms of $load \ scale$). Top row is GFM vs SM, subplot (a) has the load at the GFM bus while subplot (b) has the load at the SM bus. Bottom row has the load at bus 2, subplot (c) is SM v SM, and subplot (d) is GFM v GFM.}     

    \label{fig:2bus_stability}
\end{figure}

Fig.~\ref{fig:2bus_stability} plots the line scale at which the linearized system becomes unstable under a range of loading conditions. Four traces are plotted, one for each line model. Since the traces mostly overlap completely, the key takeaway is that all line models give very similar predictions of the line length at which instability arises. The stability `boundary' was calculated using 10 km line scale increments, so the different models may give slightly different predictions, but the differences are very small. 

There are some cases where the models deviate slightly, for example, at high load scales in Fig.~\ref{fig:inv_mach_load_inv} (GFM v SM, load at SM), and at low load scales in Fig.~\ref{fig:inv_inv} (GFM v GFM). However, considering that the total line lengths are hundreds of kilometers, these deviations are not significant.

The $MSSB$ and $MSMB$ also give almost identical results. This suggests that the extra damping associated with the $MSMB$ line does not affect the least stable eigenvalues. The closeness of these models may also reflect the limited data available for the vector fitting used to construct the $MSMB$ model.

These results are consistent with other GFM-SM simulations~\cite{markovic2021understanding} which found that the stability boundary did not differ for algebraic versus dynamic $\pi$ models. Our results extend that finding to much higher order dynamic models.  

\subsubsection{Eigenvalue analysis}

\begin{figure}
    \centering
    \includegraphics[width=.48\textwidth]{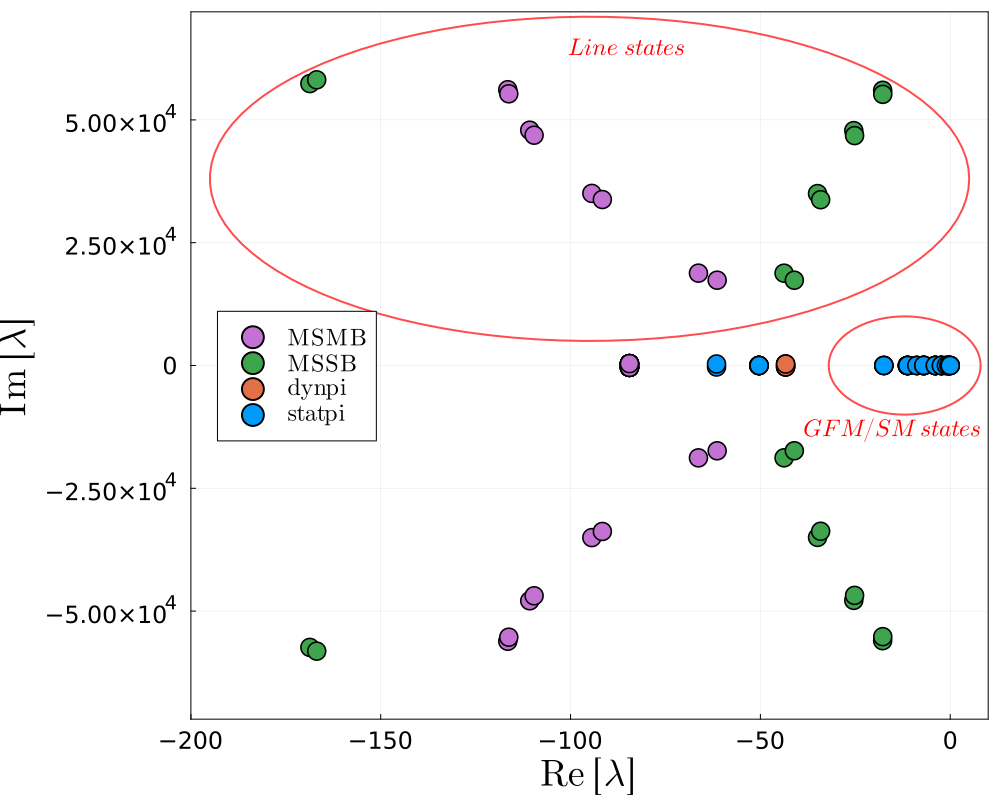}
    \caption{System eigenvalues under different line models, for a GFM v SM test case, with the load at the GFM bus, and $load\ scale = 1.0$, $line\ scale = 1.0$. }
    \label{fig:2bus_eig}
\end{figure}

Fig.~\ref{fig:2bus_eig} compares the system eigenvalues under different choices of line models for a system with nominal loading and line scaling. The blue cluster of least stable eigenvalues near the $j\omega$-axis are common to all line models, the orange eigenvalues are common to all models except $statpi$, the green to only $MSSB$, and purple to only $MSMB$. $MSSB$ (green) and $MSMB$ (purple) models exhibit higher frequency eigenvalues compared to $dynpi$, and the $MSMB$ (purple) high frequency eigenvalues are more damped compared to the $MSSB$ (green) eigenvalues.

The least stable eigenvalues, in blue, have very low ($<1\mathrm{e}^{-4}$) participation from line states. This is consistent with our finding that added line fidelity does not influence small signal stability for the cases we studied. The high frequency eigenvalues are almost exclusively associated with internal line states. The only exception is a pair of very high frequency and very damped eigenvalues, which have some small participation ($<1\mathrm{e}^{-4}$) from GFM filter voltage and SM flux states.

\begin{figure}[h]
 \begin{subfigure}{\columnwidth}
 \centering
    \includegraphics[width=\columnwidth]{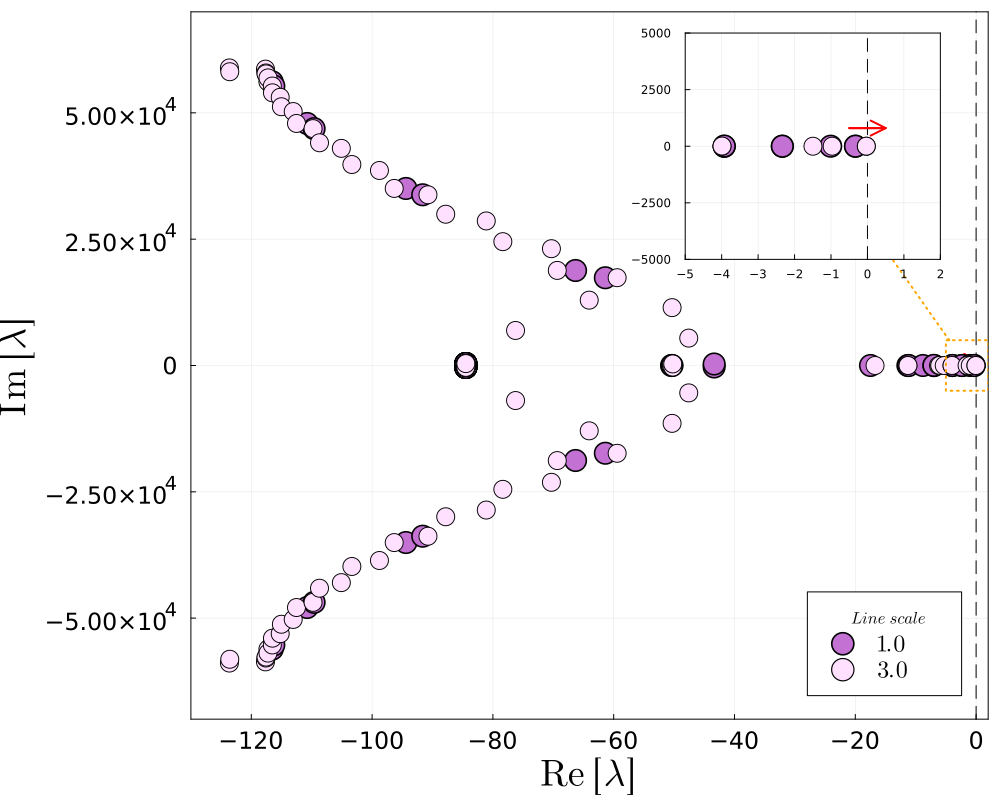}
        \caption{}
        \label{fig:line_eig_sweep}
  \end{subfigure}
    \begin{subfigure}{\columnwidth}
        \centering
        \includegraphics[width=\columnwidth]{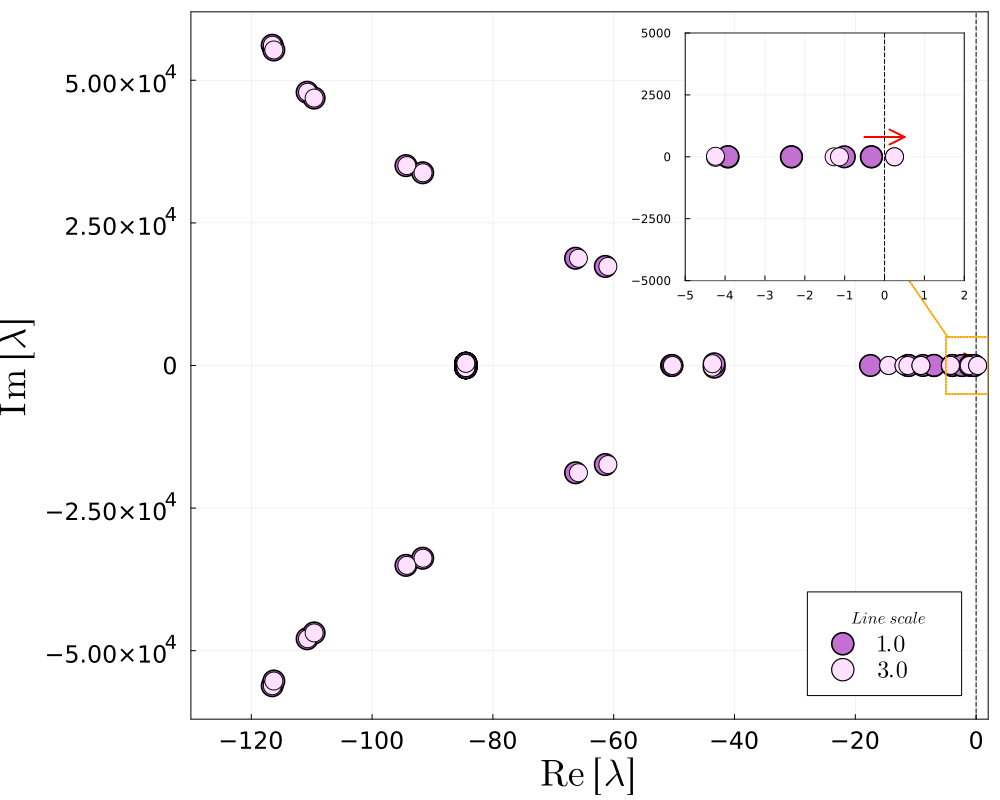}
        \caption{}
        \label{fig:load_eig_sweep}
    \end{subfigure}
    \caption{$MSMB$ eigenvalues under a sweep of (a) line lengths (with $load\ scale = 1.0$) and (b) loading (with $line\ scale = 1.0$).}
    \label{fig:eig_sweeps}
\end{figure}

In Fig.~\ref{fig:eig_sweeps} we plot how system eigenvalues move in the complex plane under changing line lengths and loading for the $MSMB$, our highest fidelity model. As we increase line lengths, we see in Fig.~\ref{fig:line_eig_sweep} that the high frequency eigenvalues (which are associated with dynamic line states) do not become less stable overall (although some eigenvalues do move right), and the cluster of least stable eigenvalues common to all dynamic line models (those with low line state participation factors) approach the right half of the complex plane (RHP) first. There is some rightward movement of high frequency line eigenvalues, and we cannot rule out the possibility that the line dynamics could destabilize the system with different parameter choices (such as control loop gains). This merits further investigation.  

In Fig.~\ref{fig:load_eig_sweep}, we see that that the high frequency line eigenvalues are very weakly influenced by loading, and instability again arises due to the cluster of common eigenvalues approaching the RHP.  

\subsection{IEEE 9 Bus test case}

The 9 bus system is stable under all load and line scalings chosen. The four line models gave very similar predictions of the maximum non-zero eigenvalue, supporting the results demonstrated on the Two Bus test case. 

\begin{figure*}[tb]
 \begin{subfigure}[t]{0.33\textwidth}
    \includegraphics[width=\textwidth]{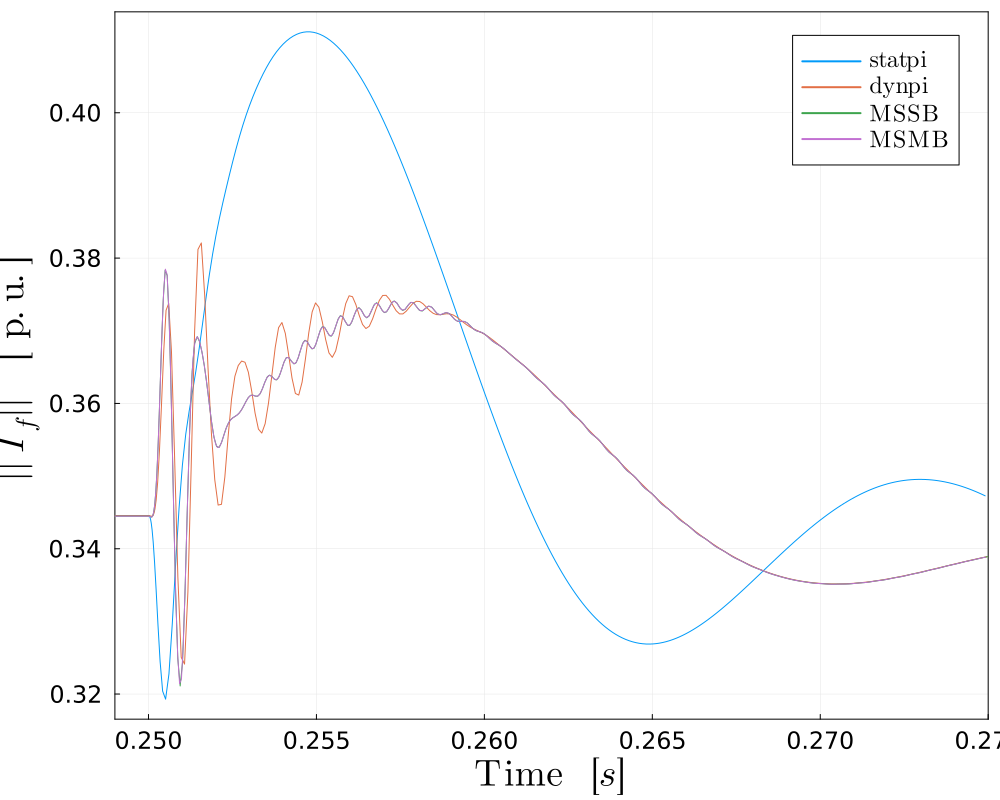}
        \label{fig:aaaa}
  \end{subfigure}
    \begin{subfigure}[t]{0.33\textwidth}
        \centering
        \includegraphics[width=\textwidth]{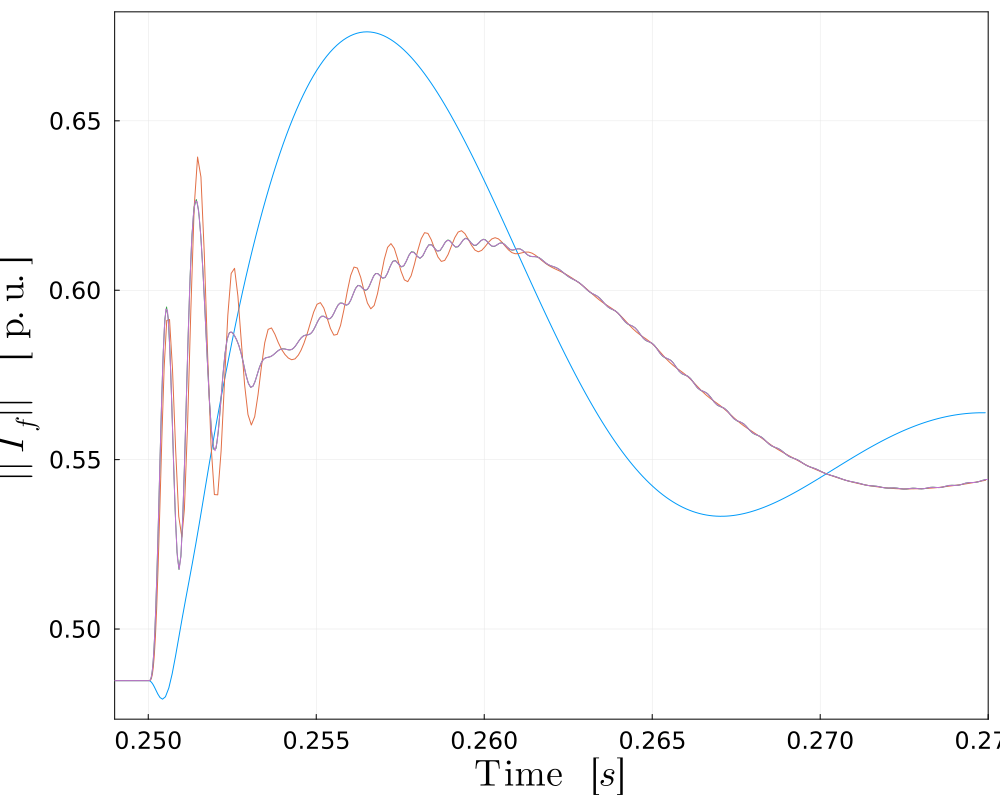}
        
        \label{fig:aaaa}
    \end{subfigure}
    \begin{subfigure}[t]{0.33\textwidth}
        \centering
        \includegraphics[width=\textwidth]{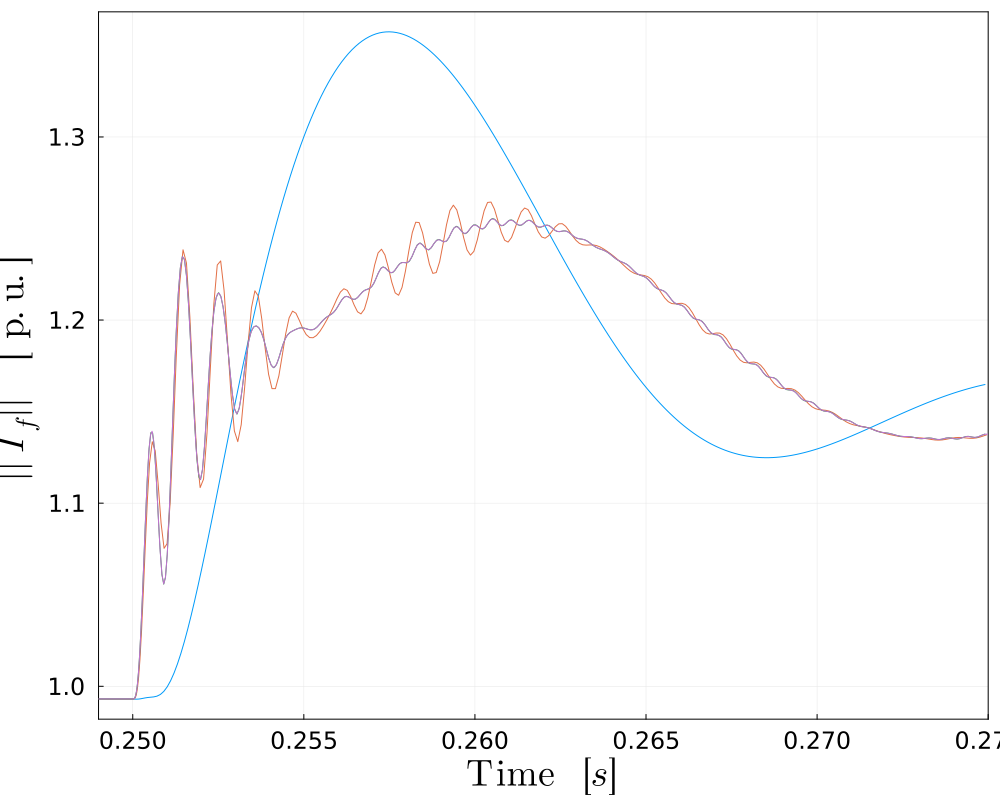}
        
        \label{fig:bbbbb}
    \end{subfigure}
    
    \label{fig:ccccc}
    \vfill
\begin{subfigure}[t]{0.33\textwidth}
    \includegraphics[width=\textwidth]{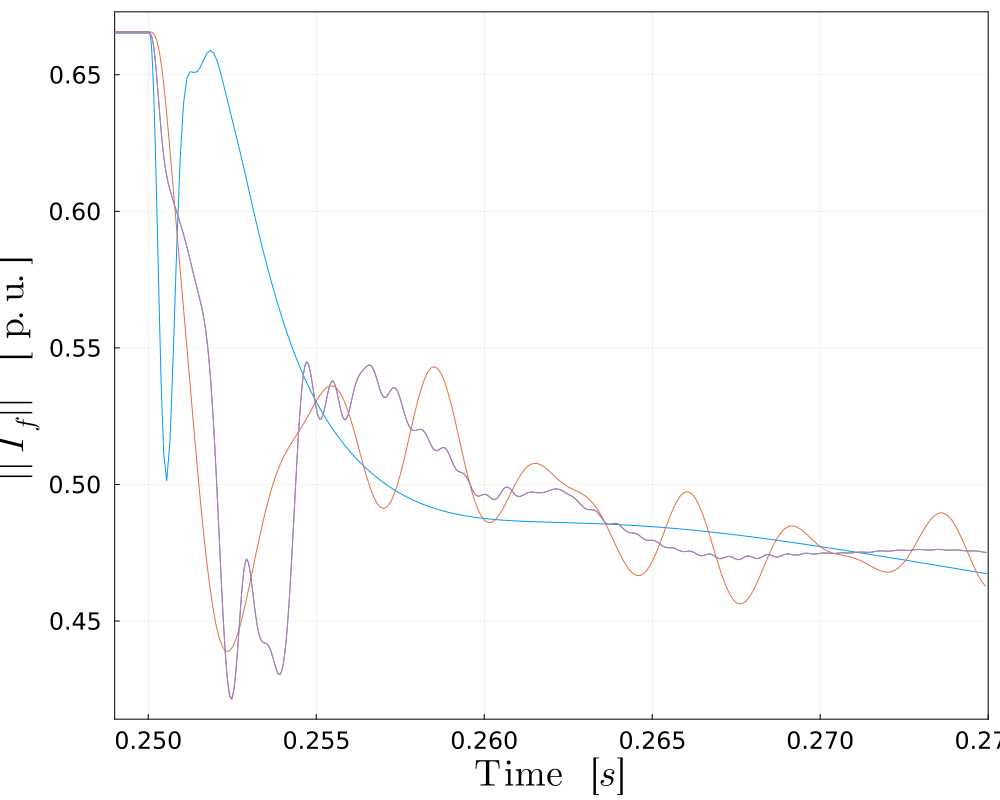}
        \caption{}
        \label{fig:aaaa}
  \end{subfigure}
    \begin{subfigure}[t]{0.33\textwidth}
        \centering
        \includegraphics[width=\textwidth]{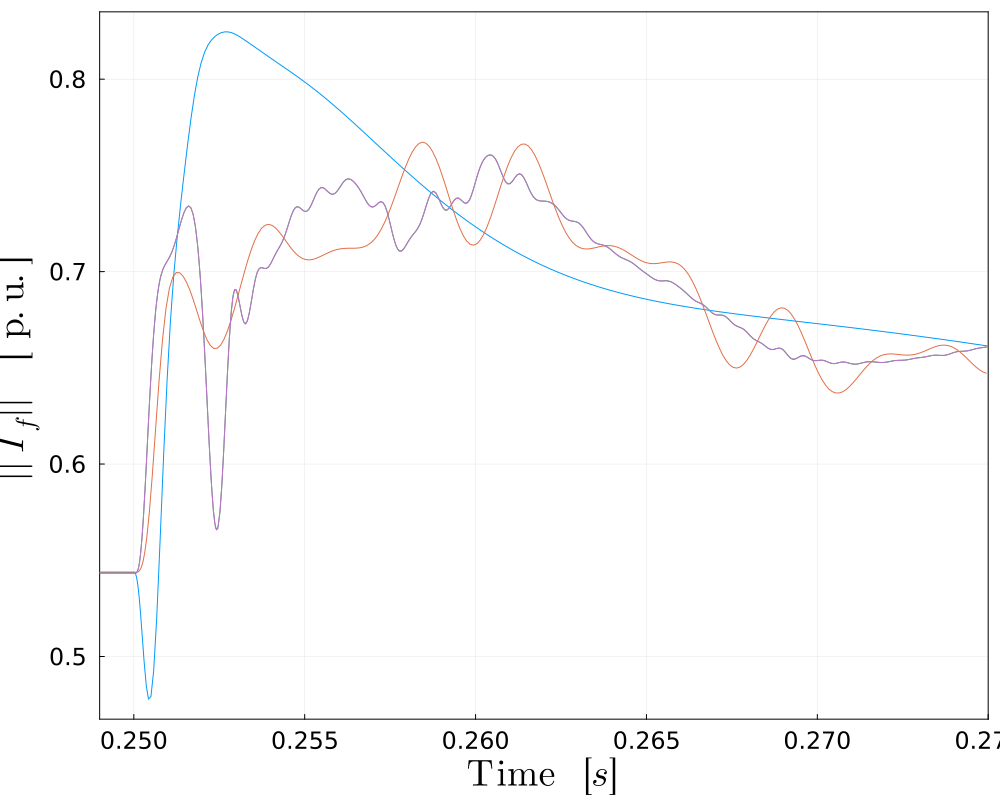}
        \caption{}
        \label{fig:aaaa}
    \end{subfigure}
    \begin{subfigure}[t]{0.33\textwidth}
        \centering
        \includegraphics[width=\textwidth]{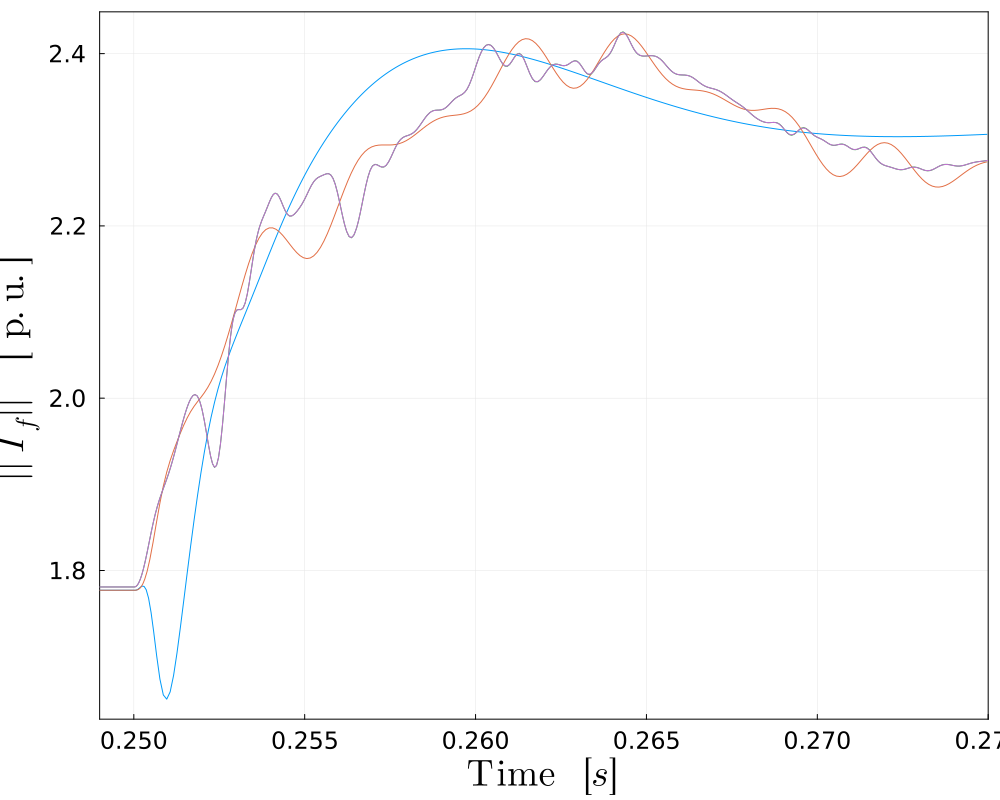}
        \caption{}
        \label{fig:bbbbb}
    \end{subfigure}
    \caption{Inverter filter current for the Two Bus test case with branch trip. The $line \ scale$ is 1.0 on the top row of plots, and 5.0 on the bottom row.  Columns (a), (b), (c) have load scaling factors of 0.5, 1.0, and 2.0, respectively.}
    \label{fig:2BusDynamics}
\end{figure*}

\begin{figure}[tb]
\includegraphics[width=\columnwidth]{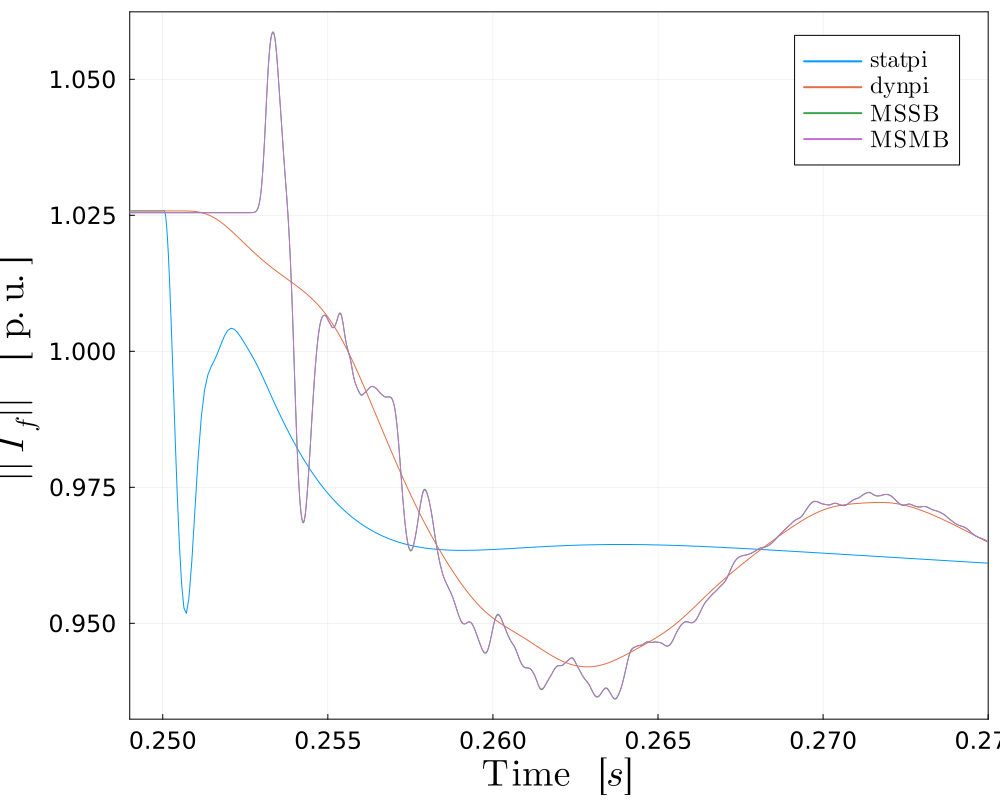}

    \caption{Bus 3 inverter filter current for the IEEE 9 Bus test case with $line \ scale = 3.0$, and $load \ scale = 1.0$. }
    \label{fig:9bus_dyn}
\end{figure}

\section{Dynamic simulations}\label{sec:dynamic_sims}
In this section, we first justify the variables that we pay close attention to, we discuss and justify the perturbation we apply to the systems, and we present the results of our dynamic studies.

\subsection{Variables, disturbance, and parameters of interest}

Inverters are known to tolerate $\sim 1.3$ p.u. rated current before risking damage to device switches.  Therefore we will focus our attention on current at inverter filters to identify whether low order line models could fail to identify inverter overcurrents.  Current dynamics can also serve as a proxy for broader system dynamics and the potential for qualitatively different simulation outcomes with different line models.  In our case we are modeling devices with LCL filters, and we therefore capture current closest to the switches.  Note that PSID does not currently model a current saturation block to protect switches from overcurrent.  Therefore our simulation results can be used to identify events in which overcurrent protection could be activated, but are  not indicative of the dynamics that would ensue following an overcurrent event. 

We chose to perturb the system with a branch trip, because it affects the system abruptly and at the differential equations closest to the variables of interest. This is in contrast to a reference change (such as in the GFM controls or SM generation setpoint), which goes through multiple filtering levels before reaching the variables of interest. A branch trip also quickly reroutes power flows throughout the network, causing fast transmission dynamics and potentially making inverters increase their current outputs to meet power demand.

As in the small signal analysis above, we test the system under a range of loading and line length parameters. 

\subsection{Two Bus test case}
For the two bus case dynamic simulations, we add the load to bus 2. We trip one of the lines since they are identical and will have the same loading. We considered several scenarios placing loads and generators in the Two Bus test case (including GFM-SM systems, GFM-only systems, and SM-only systems), with a range of loading and line lengths for each.  We find that all categories of dynamic line model produce qualitatively different dynamics following a branch trip, but that the dynamics decay to nearly the same trajectory for all models after 10 to 20 milliseconds.  Fig. ~\ref{fig:2BusDynamics} provides examples of these dynamics for the case with a SM at bus 1 and GFM plus load at bus 2.  

For the cases we investigated with the Two Bus model, we did not observe sustained line dynamics that lead to undesirable interactions between inverters and/or machines at different buses.  These results are for standard machine and CIG parameter values, but as suggested in the small signal section, we cannot exclude the possibility that undesirable interactions could occur at different parameter values (such as different CIG inner control loop gains).  Interestingly, the $MSSB$ and $MSMB$ produce essentially the same dynamics, suggesting that modeling line frequency dependence is not important for studies on this scale; this could also be a consequence of limitations of the data used for obtaining $z_{km}$ and $y_{km}$. 

We also see that, although CIG filter current oscillates more quickly with all three dynamic line models, the amplitude of these oscillations is generally smaller than, and never greater than, the amplitude of the oscillations identified with the static model.  Though there is clearly more to investigate here, these results suggest that static line models will not underestimate the potential for switch current limiting behavior.  This is likely because the distributed nature of the multi-segment lines delay and attenuate the propagation of disturbances. 



\subsection{IEEE 9 Bus test case}
In this test case we trip the line connecting buses 4 and 5 because it is the most heavily loaded line in the pre-fault system. In general, results found for the IEEE 9 Bus test case align with those of the Two Bus test case: while dynamics differ, the amplitudes of current excursions are comparable and dissipate quickly.  However, we found several instances in which $statpi$ and $dynpi$ line models underestimate inverter current magnitude excursions in important ways.  Fig.~\ref{fig:9bus_dyn} is one such example.  Here, both high-fidelity line models suggest that the initial deviation in current magnitude is not just in the opposite direction of $statpi$ and $dynpi$ lines, but its peak is significantly different than the lowest peak for the $statpi$ model. Therefore, using the $dynpi$ model may lead one to believe that the oscillations predicted by $statpi$ are not real. Hoewever, using $MSSB$ and $MSMB$ reveals that not only are oscillations present, but they occur at a different time and in a different direction from what $statpi$ predicts. This is significant in p.u. terms and suggests that high-fidelity line models should be considered when studying inverter over-current protection schemes.

\section{Conclusions and recommendations}

We find that each line model generates different sets of eigenvalues, but for the cases we investigated the dynamic line states have very low participation factors in the least stable eigenvalues, and consequently line model fidelity does not alter stability conclusions.  These results are consistent with and expand on earlier work~\cite{markovic2021understanding} that found very simple dynamic line models do not alter stability assessments in GFM-SM systems.  However, we cannot rule out the possibility that the line dynamics could destabilize the system with different parameter choices (such as control loop gains or inverter control type).  Indeed, other papers have found that simple line dynamics impact the assessment of converter gain parameters on stability~\cite{henriquez2020grid}, and that conclusions may differ with grid following converters~\cite{markovic2021understanding} or GFM devices using virtual oscillator control~\cite{Gros_Colombino_Brouillon_Dorfler_2019}.  This merits further investigation.  

Dynamic simulation results show that while $dynpi$, $MSSB$, and $MSMB$ models exhibit higher frequency components than $statpi$, these oscillations are, mostly, not of significant amplitude to be of concern and dissipate quickly. This is consistent with our small signal analysis, which indicates that dynamic line states have very low participation factors associated to eigenvalues closest to the RHP. However, we did identify cases on the IEEE 9 Bus model in which $statpi$ and $dynpi$ models fail to capture high amplitude current spikes that may be relevant to studies of inverter inner control loop performance.

For the cases we investigated, we found no evidence that modeling line frequency dependency is important. However, different line frequency datasets could produce differences in dynamic simulation results between $MSSB$ and $MSMB$, and future work should explore this possibility.

\section*{Acknowledgments}
The authors would like to acknowledge R. Henriquez-Auba's guidance in using PSID. This research was supported by the U.S. Department of Energy's Solar Energy Technologies Office through award 38637 (UNIFI Consortium).

\bibliographystyle{IEEEtran}
\bibliography{references}

\begin{thebibliography}{10}
\providecommand{\url}[1]{#1}
\csname url@samestyle\endcsname
\providecommand{\newblock}{\relax}
\providecommand{\bibinfo}[2]{#2}
\providecommand{\BIBentrySTDinterwordspacing}{\spaceskip=0pt\relax}
\providecommand{\BIBentryALTinterwordstretchfactor}{4}
\providecommand{\BIBentryALTinterwordspacing}{\spaceskip=\fontdimen2\font plus
\BIBentryALTinterwordstretchfactor\fontdimen3\font minus
  \fontdimen4\font\relax}
\providecommand{\BIBforeignlanguage}[2]{{%
\expandafter\ifx\csname l@#1\endcsname\relax
\typeout{** WARNING: IEEEtran.bst: No hyphenation pattern has been}%
\typeout{** loaded for the language `#1'. Using the pattern for}%
\typeout{** the default language instead.}%
\else
\language=\csname l@#1\endcsname
\fi
#2}}
\providecommand{\BIBdecl}{\relax}
\BIBdecl

\bibitem{ENTSOe_interaction}
``Expert group interaction studies and simulation models ({EG-ISSM}),'' Grid
  Connection European Stakeholder Committee, Tech. Rep., 2021.

\bibitem{kenyon2021validation}
R.~W. Kenyon, B.~Wang, A.~Hoke, J.~Tan, C.~Antonio, and B.-M. Hodge,
  ``Validation of maui pscad model: Motivation, methodology, and lessons
  learned,'' in \emph{2020 52nd North American Power Symposium (NAPS)}.\hskip
  1em plus 0.5em minus 0.4em\relax IEEE, 2021, pp. 1--6.

\bibitem{markovic2021understanding}
U.~Markovic, O.~Stanojev, P.~Aristidou, E.~Vrettos, D.~S. Callaway, and G.~Hug,
  ``Understanding small-signal stability of low-inertia systems,'' \emph{IEEE
  Transactions on Power Systems}, 2021.

\bibitem{henriquez2020grid}
R.~Henriquez-Auba, J.~D. Lara, C.~Roberts, and D.~S. Callaway, ``Grid forming
  inverter small signal stability: Examining role of line and voltage
  dynamics,'' in \emph{IECON 2020 The 46th Annual Conference of the IEEE
  Industrial Electronics Society}.\hskip 1em plus 0.5em minus 0.4em\relax IEEE,
  2020, pp. 4063--4068.

\bibitem{Gros_Colombino_Brouillon_Dorfler_2019}
D.~Gros, M.~Colombino, J.-S. Brouillon, and F.~Dorfler, ``The effect of
  transmission-line dynamics on grid-forming dispatchable virtual oscillator
  control,'' \emph{IEEE Transactions on Control of Network Systems}, vol.~6,
  no.~3, p. 1148–1160, Sep 2019.

\bibitem{D’Arco_Beerten_Suul_2015_cable_MOR}
\BIBentryALTinterwordspacing
S.~D’Arco, J.~Beerten, and J.~Suul, ``\BIBforeignlanguage{en}{Cable model
  order reduction for hvdc systems interoperability analysis},'' in
  \emph{\BIBforeignlanguage{en}{11th IET International Conference on AC and DC
  Power Transmission}}.\hskip 1em plus 0.5em minus 0.4em\relax Birmingham, UK:
  Institution of Engineering and Technology, 2015, pp. 026 (10 .)--026 (10 .).
  [Online]. Available:
  \url{https://digital-library.theiet.org/content/conferences/10.1049/cp.2015.0039}
\BIBentrySTDinterwordspacing

\bibitem{D’Arco_Suul_Beerten_2019}
\BIBentryALTinterwordspacing
S.~D’Arco, J.~A. Suul, and J.~Beerten, ``Time-invariant state-space model of
  an ac cable by $dq$-representation of frequency-dependent $\pi$-sections,''
  in \emph{2019 IEEE PES Innovative Smart Grid Technologies Europe
  (ISGT-Europe)}.\hskip 1em plus 0.5em minus 0.4em\relax Bucharest, Romania:
  IEEE, Sep 2019, p. 1–5. [Online]. Available:
  \url{https://ieeexplore.ieee.org/document/8905577/}
\BIBentrySTDinterwordspacing

\bibitem{D’Arco_Suul_Beerten_2021_config}
------, ``Configuration and model order selection of frequency-dependent $\pi$
  models for representing dc cables in small-signal eigenvalue analysis of hvdc
  transmission systems,'' \emph{IEEE Journal of Emerging and Selected Topics in
  Power Electronics}, vol.~9, no.~2, p. 2410–2426, Apr 2021.

\bibitem{lara2023powersimulationsdynamicsjl}
J.~D. Lara, R.~Henriquez-Auba, M.~Bossart, D.~S. Callaway, and C.~Barrows,
  ``Powersimulationsdynamics.jl -- an open source modeling package for modern
  power systems with inverter-based resources,'' 2023.

\bibitem{revisiting_methods_lara}
J.~D. Lara, R.~Henriquez-Auba, D.~Ramasubramanian, S.~Dhople, D.~S. Callaway,
  and S.~Sanders, ``Revisiting power systems time-domain simulation methods and
  models,'' \emph{IEEE Transactions on Power Systems}, pp. 1--16, 2023.

\bibitem{Beerten_D’Arco_Suul_2016}
J.~Beerten, S.~D’Arco, and J.~A. Suul,
  ``\BIBforeignlanguage{en}{Frequency‐dependent cable modelling for
  small‐signal stability analysis of vsc‐hvdc systems},''
  \emph{\BIBforeignlanguage{en}{IET Generation, Transmission \& Distribution}},
  vol.~10, no.~6, p. 1370–1381, Apr 2016.

\bibitem{milano2010power}
F.~Milano, \emph{Power system modelling and scripting}.\hskip 1em plus 0.5em
  minus 0.4em\relax Springer Science \& Business Media, 2010.

\bibitem{yazdani2010voltage}
A.~Yazdani and R.~Iravani, \emph{Voltage-sourced converters in power systems:
  modeling, control, and applications}.\hskip 1em plus 0.5em minus 0.4em\relax
  John Wiley \& Sons, 2010.

\bibitem{kenyon2021open}
R.~W. Kenyon, A.~Sajadi, A.~Hoke, and B.-M. Hodge, ``Open-source pscad
  grid-following and grid-forming inverters and a benchmark for zero-inertia
  power system simulations,'' in \emph{2021 IEEE Kansas Power and Energy
  Conference (KPEC)}.\hskip 1em plus 0.5em minus 0.4em\relax IEEE, 2021, pp.
  1--6.

\bibitem{gustavsen1999rational}
B.~Gustavsen and A.~Semlyen, ``Rational approximation of frequency domain
  responses by vector fitting,'' \emph{IEEE Transactions on power delivery},
  vol.~14, no.~3, pp. 1052--1061, 1999.

\bibitem{Dommel_1985}
H.~Dommel, ``Overhead line parameters from handbook formulas and computer
  programs,'' \emph{IEEE Transactions on Power Apparatus and Systems}, vol.
  PAS-104, no.~2, p. 366–372, Feb. 1985.

\bibitem{9bus_figure}
Y.~Song, D.~Hill, and T.~Liu, ``Small-disturbance angle stability analysis of
  microgrids: A graph theory viewpoint,'' 09 2015.

\bibitem{Berard2017}
J.~Bérard, ``Ieee 9 bus system example,'' 2017.

\end{thebibliography}
\end{document}